\documentclass[prl,twocolumn, superscriptaddress,floatfix,showpacs]{revtex4}

\usepackage{graphicx}
\usepackage{amsbsy,amssymb,amsmath,bm}
\usepackage{amsfonts}
\usepackage{textcomp}
\usepackage{epsfig}

\begin{document}

\preprint{}

\title{Direct evidence for the suppression of charge stripes in epitaxial La$_{1.67}$Sr$_{0.33}$NiO$_4$ thin films}

\author{C. K. Xie}
\author{J. I. Budnick}
\author{W. A. Hines}
\author{B. O. Wells}%
\affiliation{Department of Physics, University of Connecticut, Storrs, CT 06269}%

\author{Feizhou He}
 \affiliation{Canadian Light Source, University of Saskatchewan, Saskatoon, SK S7N 0X4, Canada}%

\author{A. R. Moodenbaugh}
 \affiliation{Condensed Matter Physics and Materials Science Department, Brookhaven National Laboratory, Upton, NY 11973-5000}%

\date{\today}

\begin{abstract}
We have successfully grown epitaxial La$_{1.67}$Sr$_{0.33}$NiO$_4$
 films with a small crystalline mosaic using pulsed laser deposition. With synchrotron
radiation, the x-ray diffraction peaks associated with charge
stripes have been successfully observed for relatively thick films.
Anomalies due to the charge-ordering transition have been examined
using four-point probe resistivity measurements. X-ray scattering
provides direct evidence for suppression of the stripe phase in thin
samples; the phase disappears for film thicknesses $\leqslant$ 2600
~\AA{}. The suppression appears to be a result of shrinking the
stripe phase domains. This may reflect the stripe phase progressing
from nematic to isotropic.

\end{abstract}

\maketitle

The real space ordering of charges, spins, and electronic orbitals
in correlated electron materials has been a major topic in condensed
matter physics in recent years. Such behavior has been found in a
variety of transition metal oxides that exhibit a range of physical
properties. A typical situation occurs when antiferromagnetic
insulators of these oxides are electronically doped, the charge
carriers tend to localize and order ~\cite{Cc93,Jt95}. One of the
most important types of order is generally known as stripes. This is
best known in doped compounds of the La$_2$CuO$_4$ family of high
temperature superconductors. At a particular density of doped holes,
$n_{h} = 1/8$, there is a static, ordered arrangement of spins and
charges forming stripes along the Cu-O bond direction~\cite{Jt97}.
The isostructural system La$_{2-x}$Sr$_{x}$NiO$_{4+y}$ shows a
similar stripe structure but with an orientation rotated by
45\ensuremath{^\circ} with respect to the Ni-O bond direction in the
planes. However, it does not exhibit superconductivity and even
remains insulating for a wide doping range ~\cite{Jt94}.
Furthermore, a variety of manganese oxide materials appears to show
various types of charge, spin, and orbital ordering, some of which
are stripe-like ~\cite{sm98, pd00}. More recently, a short-range
charge ordering in Ho-doped SrCoO$_{3-x}$ cobaltite with
charge-ordered clusters size down to 50~\AA{} is observed for a
broad compositional range~\cite{str06}. As more examples of charge
inhomogeneity are discovered, one might think that the spin-charge
ordering may be a ubiquitous property of transitional metal oxides.
However, there are certainly many examples of charge-doped
transition metal oxides where no such charge inhomogeneity has been
reported. It is thus important to understand the nature of charge
ordering correlations, their effect on physical properties, and
under what situations they will occur. To further such a goal, it
would be very helpful to find a single materials system where the
charge orders through an external tuning parameter that does not
affect the charge concentration itself.

There have been several efforts to investigate the correlation
between the stripe phase and lattice distortion by applying
mechanical strain in cuprate materials. The static stripe phase in
La$_{1.875}$Ba$_{0.125}$CuO$_4$ and
(LaNd)$_{1.875}$Sr$_{0.125}$CuO$_4$ is accompanied by a structural
phase transition from a low temperature orthorhombic (LTO) to a low
temperature tetragonal (LTT) phase and an anomalous suppression of
superconductivity~\cite{Ma88, Na98}. It appears that the application
of hydrostatic pressure to materials with the 1/8 doped stripe phase
causes the LTO to LTT structural phase transition to be suppressed
and superconductivity to be partially recovered, with a transition
temperature (T$_C$) of 15 K.~\cite{Sk93,arnie1,Mk05,Sa02} Studies of
similar cuprate films with an in-plane compressive strain have shown
a similar effect~\cite{Yl04, Ys00}. However, since none of these
studies provide direct information about the behavior of stripe
phase, their interpretation may be questioned. Typically, it has
been assumed that the static charge stripe phase requires the
presence of the LTT structural phase. However, it is possible that
the stripe phase can form without the structural phase transition
and indeed some recent studies have found just such a
case~\cite{wells}. Therefore, direct evidence is important to
understand the nature of the stripe phase under these conditions.

In this letter, we report studies of the behavior of the stripe
phase in La$_{1.67}$Sr$_{0.33}$NiO$_4$ (LSNO) films as a function of
film thickness. The nickelate films were chosen for study because
the appropriate stoichiometry is relatively stable and the stripe
ordering is pronounced, with relatively strong diffraction peaks and
a high stripe ordering temperature of 240 K. Using direct evidence
from x-ray diffraction (XRD) and resistivity measurements, we
demonstrate that the stripe phase in LSNO films is suppressed as the
films become thin. By examining the scattering profiles and the full
temperature dependent resistivity we can understand how the stripes
disappear and what short range order remains.

Epitaxial LSNO films were deposited on SrTiO$_3$ (STO) substrates
using a pulsed laser deposition technique. The substrate temperature
was kept at 780 \textcelsius~ in 100 mTorr oxygen during the
deposition. After the deposition, the samples were slowly cooled
down to room temperature in 200 Torr oxygen. The thicknesses of the
LSNO films range from 480 \AA{} to 8000 \AA{}. X-ray diffraction
showed excellent epitaxy with mosaics ranging from
0.27\ensuremath{^\circ} to 0.36\ensuremath{^\circ} and no detectable
disoriented regions. XRD measurements were carried out at beamline
X22A and X22C at the National Synchrotron Light Source, Brookhaven
National Laboratory. A graphite (002) single bounce analyzer was
used for suppressing the background counts and eliminate high order
harmonics to reveal the charge stripe peaks. Finite thickness
oscillations were observed in reflectometry measurements, indicating
a smooth surface and also giving a measurement of the film
thickness. The samples were cooled in a closed-cycle refrigerator
between 15 K $\leqslant$ T $\leqslant $ 300 K with a temperature
control better than 0.5 K. A standard four-probe technique was
utilized to obtain the electrical transport properties of the films
as a function of temperature ranging from 140 K to 340 K in a
Quantum Design MPMS system equipped with a Manual Insertion Utility
Probe and External Device Control software package.

The stripe phase in La$_{2-x}$Sr$_{x}$NiO$_4$ can be studied through
the examination of magnetic and nuclear superlattice peaks observed
in diffraction experiments~\cite{Tr1}. With synchrotron XRD, we are
able to directly study the charge stripe phase in films by the
inspection of incommensurate supperlattice peaks associated with the
charge stripe order. These peaks in diffraction techniques result
from a slight periodic lattice distortion in response to the ordered
inhomogenity of doped holes, and are typically 10$^{8}$ weaker in
magnitude than a Bragg peak. We have successfully observed the
charge ordering peaks in relatively thick LSNO films. Fig. 1
presents linear scans over the stripe peaks (0.67, 0, 9) and (1.33,
0, 9) of an 8000~\AA{} LSNO film on a STO substrate along H, K, and
L directions in reciprocal space, respectively. The observed peaks
are well developed at a low measurement temperature of about 15 K.
The solid lines represent the best fit with a Lorentzian function,
from which the intensity and peak width can be derived. The width of
the charge stripe peaks in the H and K directions are approximately
equal, with the full width at half maximum (FWHM) of 0.049 and 0.033
reciprocal lattice units (r.l.u.), respectively. However, the peak
is much broader along the L direction, with FWHM of 0.28 r.l.u. The
inverse correlation length $\xi^{-1}$ is defined as $\xi_{d}^{-1} =
\frac{2\pi}{d} \emph{w}$, where $\emph{w}$ is the half width at half
maximum (HWHM) of the reflection and \emph{d} is the lattice
constant. According to this, the charge order has a correlation
length of approximately 7 and 10 unit cells in the H and K
direction, respectively. The correlation length is only 1 unit cell
in the L direction. This suggests that the charge ordering is
primarily two dimensional in nature. Our result is in agreement with
that from a bulk LSNO crystal sample, though the correlation lengths
are smaller in our films~\cite{Du}.

\begin{figure}
\includegraphics[scale=.6]{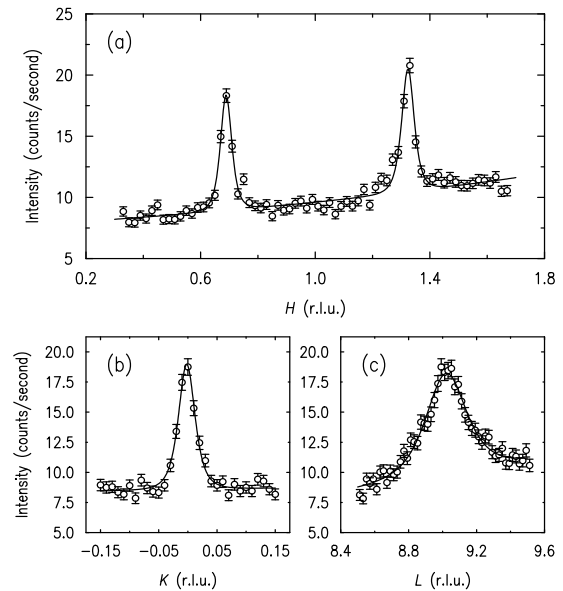}

\caption{\label{fig:Fig1.eps} The charge stripe peak of a 8000\AA{}
LSNO film on a STO substrate at 15 K. (a) H scan along (\emph{h}, 0,
9), showing two peaks at \emph{h} =0.67 and 1.33; (b) K scan along
(0.67, \emph{k}, 9); (c) L scan along (0.67, 0, \emph{l} ) }
\end{figure}

\begin{figure}

\includegraphics[scale=.6]{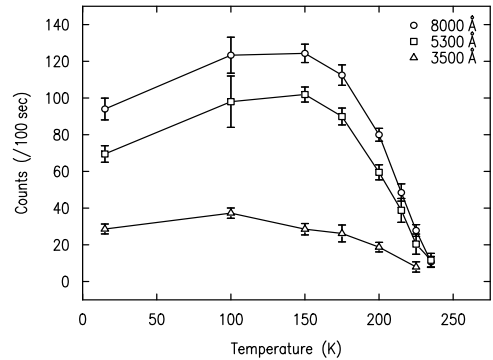}
\caption{\label{fig:Fig2.eps} Intensity of the stripe peak (0.67, 0,
9) versus temperature for LSNO films with various thicknesses.}
\end{figure}

The intensity of the charge stripe peak (0.67, 0, 9) as a function
of temperature is shown in Fig. 2, for a series of LSNO films on STO
substrates with different thicknesses. The stripe ordering
temperature is about 240 K for all films, in agreement with
experiments on bulk materials~\cite{hatton}. It appears that the
ordering temperature does not change with the film thickness. Upon
cooling, the peak intensity increases steadily and reaches a maximum
at about 140 K and then decreases slightly for lower temperatures.
The loss of intensity at low temperatures appears to be universal
for all our LSNO films. This phenomenon is also seen in bulk
crystals when investigated using x-ray diffraction~\cite{Me04}.

Fig. 3 shows the thickness dependence of the integrated intensity of
stripe peaks (0.67, 0, 9) and their width, all at 100 K. The
integrated intensity decreases in a roughly linearly manner, as we
would expect from a simple scattering volume argument for
progressively thinner films. Thus the disappearance of the stripe
peak is primarily through the broadening of the peaks. For samples
where the stripe peaks can be detected, the peak intensity per
scattering volume remains roughly constant and there is very little
change in the stripe ordering temperature. The broadening of the
stripe peak width suggests that the domain size of the stripe phase
shrinks as the film thickness decreases. An examination of the (2,
0, 8) Bragg peak for the same films suggests that the main
structural peaks are much narrower, about 0.01 r.l.u., and more
importantly do not broaden as the films become thin. Thus, the
crystallinity of the films themselves is similar for the different
thicknesses and we can attribute the increased width of the charge
order peak directly to domain size. Compared with the 8000~\AA{}
thick film, the stripe peak from the thin film, 3500~\AA{}, has
broadened such that the correlation length of the stripe phase has
decreased by about $50\%$ to nearly 3 unit cells. For thinner films,
we could not observe any stripe peak, suggesting further reduction
of the correlation length. Thus the charge stripes disappear as
their domain size is reduced to levels below which stripes cannot be
measured and stripe order may no longer be meaningful.

The films which show stripe order disappearing in Fig. 3 do not have
a significant strain as they are all thick enough to be fully
relaxed. Thus the suppression of stripe order is not a direct effect
of strain or microscopic parameters that depend on bond length.
However, these films are epitaxially constrained to remain in
registry with the substrate. This clamping effect will tend to
suppress any transition with a structural reconstruction. The effect
has been known to alter structural phase transitions in various ways
~\cite{He}. The energy cost of any interface mismatch varies with
the area of the interface, but energy cost of suppressing the bulk
transition varies with the entire film volume. Thus we expect the
clamping effect to be more important for thinner films but still be
fairly long range. The elastic effects of epitaxy increase the free
energy for any phase on the low temperature side of a phase
transition with a structural component and thus may be expected to
suppress the phase transition. This result uniquely demonstrates
that the stripe phase can be switched off through the influence of a
structural control parameter, suggestive of the sensitivity of the
stripes to the lattice distortion~\cite{Ma88}.

\begin{figure}
\centering
\epsfig{file=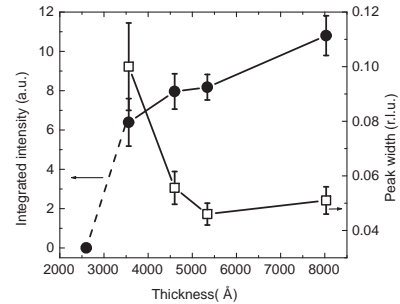,width=6cm,clip=}
\caption{\label{fig:Fig3.eps} Integrated intensity and peak width of
the stripe phase (0.67, 0, 9) versus film thickness. The dashed line
shows a sudden drop of the peak intensity, which indicates the
suppression of the stripe phase. All LSNO films are measured at 100
K.}
\end{figure}

We also studied the temperature dependence of the resistivity of
LSNO films. The first derivative of resistivity versus temperature
measurements for a thick film (4600~\AA{}) and a thin film
(480~\AA{}) on STO(100) substrate are displayed in Fig. 4(a) and
Fig. 4(b), respectively. The thick sample was measured to have
stripes as shown in Fig. 3, the thinner sample did not. The insets
of Fig. 4 show the resistivity as a function of temperature.
Insulating behavior is demonstrated in both films. For the thick
sample, there is a perceptible kink in the slope (Fig. 4(a)). This
becomes clearer in the derivative curve, with a minimum near 220 K.
This kink appears to be associated with the charge ordering
transition, and has been identified as such in transport
measurements conducted on bulk samples~\cite{Sw94}. The anomaly in
transport properties is thus indicative of the formation of charge
stripes in the film. As shown in Fig. 4(b), the development of the
resistivity as a function of temperature for thin films with no
stripe ordering is smooth, without a dip in the derivative curve.
Unlike 1/8-doped cuprates, there is no structural phase transition
for LSNO that is related to the formation of the stripe
phase~\cite{Hucker}. Thus this anomaly in resistivity is not related
to any change in atomic structure, such as the transition from LTO
to LTT in cuprates, but must be due to the ordering of doped charges
themselves. The absence of the anomaly in the resistivity confirms
that the stripe phase itself is truly suppressed in thin films. The
transport measurements are consistent with the stripe phase
disappearing through a collapse of the correlation length or domain
size. Away from the transition, the actual resistivity values for
the thick and thin films are similar. We would expect little change
in parameters such as the hopping energy of the doped holes if in
fact the local order is unaffected and only longer range ordering is
suppressed. Therefore, at low temperatures the stripe-like order of
the doped holes is not simply gone. The stripe phase is suppressed
as the size of the stripe domains collapses. However, the local
hopping environment for the holes remains essentially unchanged.

\begin{figure}
\centering
\epsfig{file=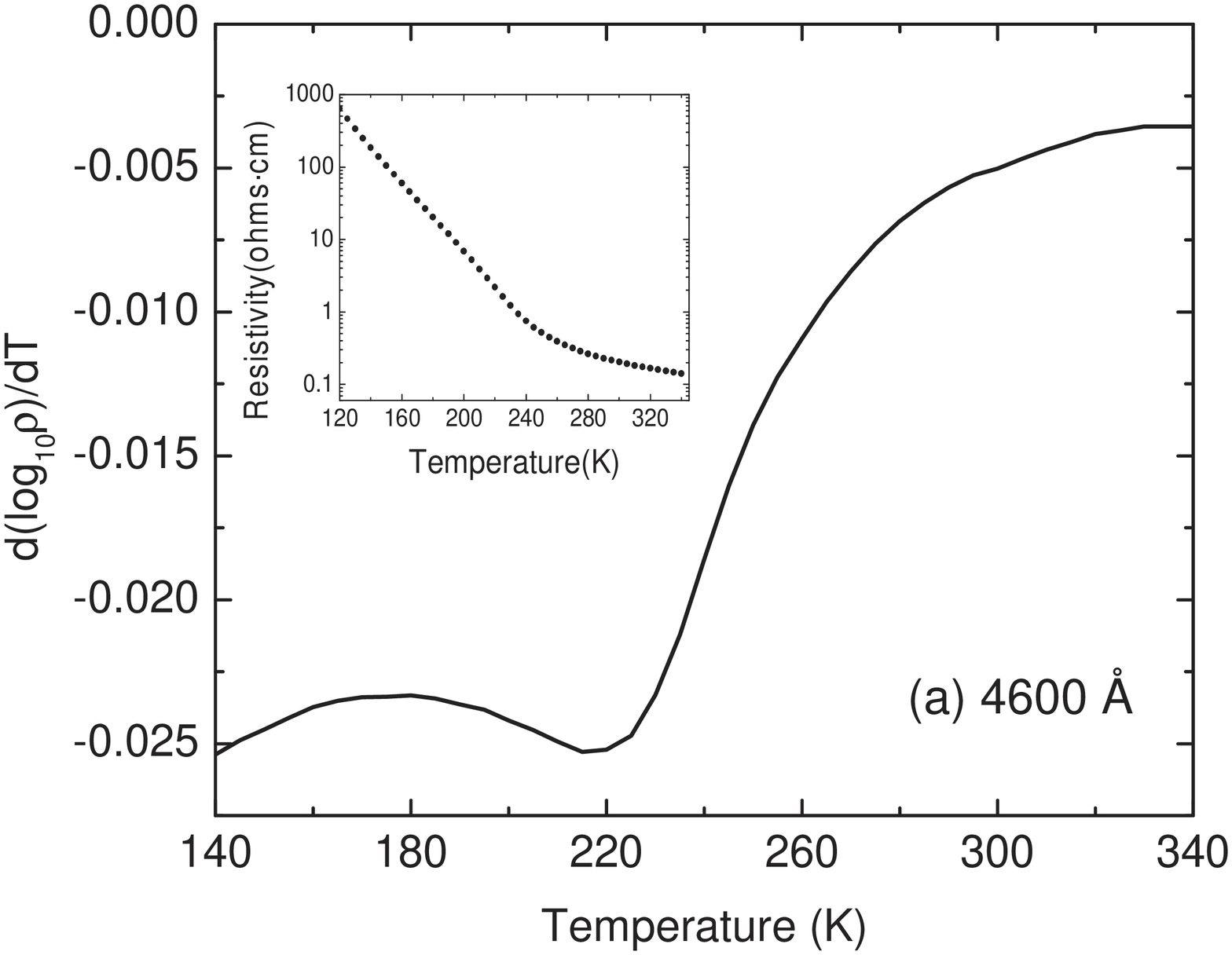,width=6cm,clip=},
\epsfig{file=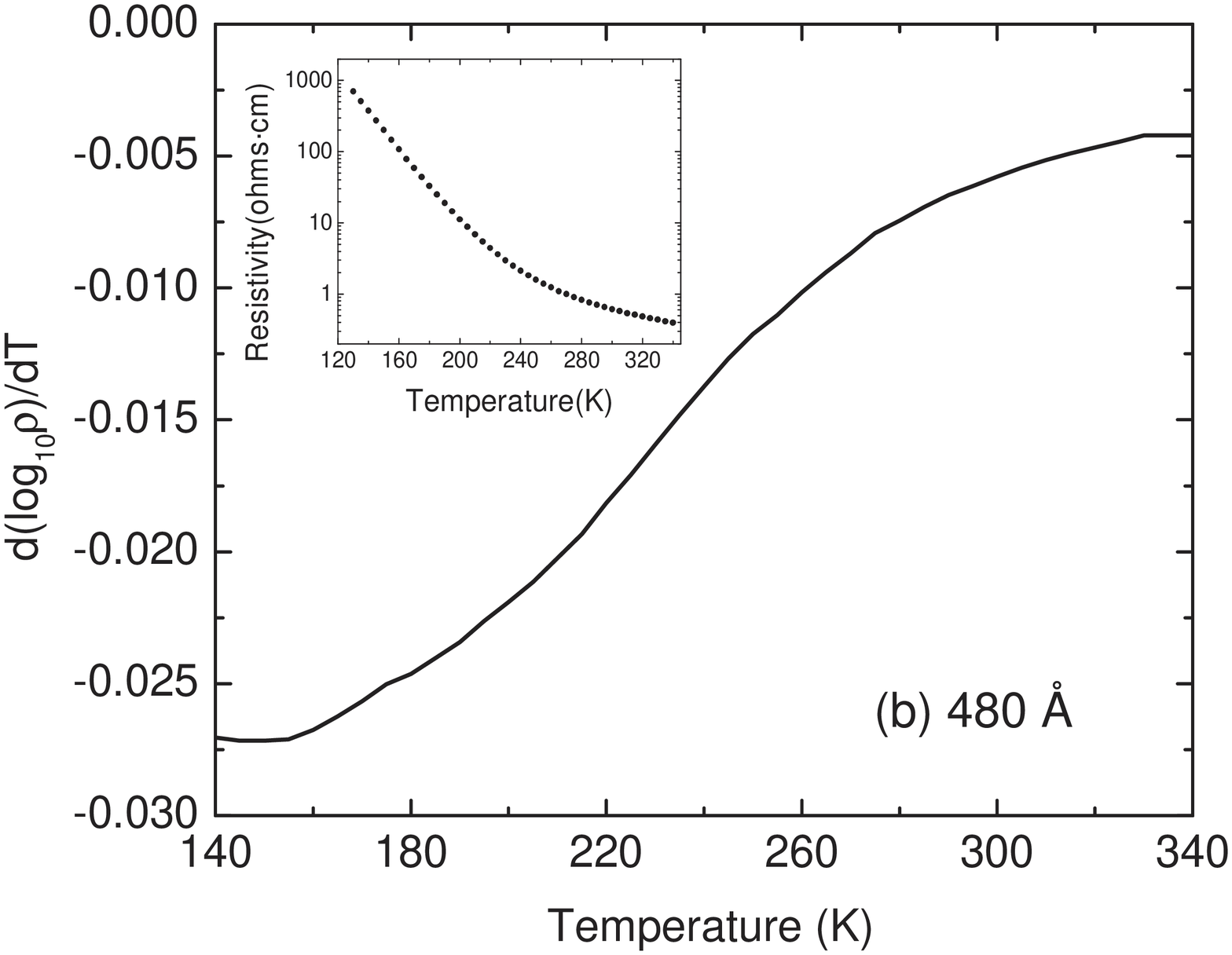,width=6cm,clip=} \\
\caption{\label{fig:Fig4.eps}Resistivity versus temperature for LSNO
films. (a) film thickness 4600\AA{} (b) film thickness 480\AA{}. The
480\AA{} film does not show the resistivity anomaly.}
\end{figure}

A popular model to describe the stripe-like charge ordering has been that of an electronic liquid
crystal ~\cite{Kivelson}~\cite{Zaanen}. Within this model there is a direct analogy between the
various stripe-like features found in the layered transition metal oxides and the crystalline,
smectic, nematic, and isotropic phases of classical liquid crystals. According to Kivelson et
al.~\cite{Kivelson}, it appears that the stripes we measure are most likely in a nematic or a
disrupted smectic phase. The diffraction profile for the thicker films is much like that found in
(LaNd)$_{2-x}$Sr$_x$CuO$_4$ but with considerably broader peaks in the in-plane directions. Such a
profile  may arise from a disrupted smectic phase with significant disorder as one might expect in
a film. However, the peaks further broaden as the film thickness decreases while the film main
Bragg peaks remain sharp. The broadening of the charge order peaks as a function of decreasing film
thickness is qualitatively similar to the broadening seen as a function of temperature in classical
liquid crystals~\cite{chaikin}. When the scattering peaks disappear for very thin films, the
stripes may be entering an isotropic or may still be nematic with peaks too broad to measure here.
The fact that the low temperature resistivity has not changed substantially would indicate that the
stripes maintain their local integrity in the samples measured. It is of great interest that the
changes induced through the film-substrate interactions as the films become thinner appear to be
opposite to those of a magnetic field on bulk samples of the cuprates. Neutron diffraction has
revealed that the magnetic peak associated with stripe-like ordering grows in intensity with an
applied magnetic field~\cite{Lake}. That was explained as a nematic to smectic transition by Zaanen
et al.~\cite{Zaanen}. While the transition studied here moves in the opposite direction, nematic to
isotropic, both data sets are characterized by a transition temperature that does not change while
the intensity is altered by the tuning parameter.

Finally, we note an apparent similarity between the topography of
stripes we infer and that measured in recent scanning tunneling
microscopy (STM) studies~\cite{Kohsaka}. Diffraction has been the
main tool for identifying charge-ordered phases, striped phases for
layered perovskite oxides. However, only a limited number of
materials have shown appropriate diffraction peaks. The clearest
cases are La$_{2-x}$Sr$_x$NiO$_4$~\cite{Hucker} and the 1/8th doped
cuprates~\cite{Ma88,Na98}. Another possible candidate for stripe
phases identified through diffraction is superoxygenated and phase
separated La$_{2-x}$Sr$_x$CuO$_{4+y}$~\cite{wells}. Some other
cuprates have shown magnetic diffraction consistent with striped
phases but without the charge order peaks~\cite{katano,Lake}. Our
work shows that even when diffraction does not detect an ordered
striped phase, and where resistivity does not detect a transition,
the layered nickelates may still support at least an incipient form
of electronic stripes. This might be similar to the phenomenon seen
in underdoped cuprates~\cite{Kohsaka}, where STM images revealed
features of stripe phase but with short range order correlation
length $\sim$ 4a$_0$$\times$ 4a$_0$, where the a$_0$ is the lattice
constant of the sample. Those samples are not known to show
diffraction peaks associated with stripes, but are in a similar
doping region to samples that do show stripes. Taken together, this
is evidence that at least an incipient form of electronic stripes
might be more prevalent in layered transition metal oxides than
previously believed. While measurement of non-ordered phases is
difficult, it would be valuable to investigate a wider range of
layered, transition metal oxides for possible short-ranged stripe
order.

This work is supported through NSF DMR-0239667, and the USDOE,
DE-FG02-00ER45801. Work at Brookhaven National Laboratory, is
supported by the U.S. Department of Energy under contract
DE-AC02-98CH1-886, at CMPMSD by Division of Materials Sciences, at
NSLS additionally by Division of Chemical Sciences. B.O.W.
acknowledges support from the Cottrell Scholar program of the
Research Corporation.


\end{document}